\documentclass[english,twocolumn,prl,amsmath,amssymb,floatfix,superscriptaddress]{revtex4-2}
\usepackage{dcolumn}
\usepackage{bm}
\usepackage{graphicx}
\usepackage{color}
\usepackage{babel}
\usepackage{amsfonts}
\usepackage{mathtools}
\usepackage{slashed}
\usepackage{enumerate}
\usepackage{braket}
\usepackage[multiple]{footmisc}

\usepackage{tikz}
\usetikzlibrary{calc,shapes.arrows,shapes.geometric}
\newcommand{\be}{\begin{equation}}
\newcommand{\ee}{\end{equation}}
\newcommand{\bea}{\begin{eqnarray}}
\newcommand{\eea}{\end{eqnarray}}

\begin{document}


\title{Fractional entropy of multichannel Kondo systems from conductance-charge relations}


\author{Cheolhee Han}
\affiliation{Raymond and Beverly Sackler School of Physics and Astronomy, Tel Aviv University, Tel Aviv 69978, Israel}
\author{Z. Iftikhar}
\affiliation{Universit\'e Paris-Saclay, CNRS, Centre de Nanosciences et de Nanotechnologies (C2N), 91120 Palaiseau, France}
\author{Yaakov Kleeorin}
\affiliation{Center for the Physics of Evolving Systems, 
University of Chicago , Chicago, IL, 60637, USA}
\author{A. Anthore}
\affiliation{Universit\'e Paris-Saclay, CNRS, Centre de Nanosciences et de Nanotechnologies (C2N), 91120 Palaiseau, France}
\affiliation{Universit\'{e} de Paris,  F-75006 Paris, France}
\author{F. Pierre}
\affiliation{Universit\'e Paris-Saclay, CNRS, Centre de Nanosciences et de Nanotechnologies (C2N), 91120 Palaiseau, France}
\author{Yigal Meir}
\affiliation{Department of Physics, Ben-Gurion University of the Negev, Beer-Sheva, 84105 Israel}
\author{Andrew K. Mitchell}
\email[]{andrew.mitchell@ucd.ie}
\affiliation{School of Physics, University College Dublin, Belfield, Dublin 4, Ireland}
\affiliation{Centre for Quantum Engineering, Science, and Technology, University College Dublin,  Dublin 4, Ireland}
\author{Eran Sela}
\email[]{eranst@post.tau.ac.il}
\affiliation{Raymond and Beverly Sackler School of Physics and Astronomy, Tel Aviv University, Tel Aviv 69978, Israel}


\begin{abstract}
Fractional entropy is a signature of nonlocal degrees of freedom, such as Majorana zero modes or more exotic non-Abelian anyons.
Although direct experimental measurements remain challenging, Maxwell relations provide an indirect route to the entropy through charge measurements. 
Here we consider multichannel charge-Kondo systems, which are predicted to host exotic quasiparticles due to a frustration of Kondo screening at low temperatures. 
In the absence of experimental data for the charge occupation, we derive relations connecting the latter to the conductance, for which experimental results have recently been obtained. Our analysis indicates that Majorana and Fibonacci anyon quasiparticles are well-developed in existing two- and three-channel charge-Kondo devices, and that their characteristic  $k_{\rm{B}}\log\sqrt{2}$ and $k_{\rm{B}}\log\frac{1+\sqrt{5}}{2}$ entropies are experimentally measurable. 
\end{abstract}
\maketitle


\noindent A plethora of condensed-matter systems are conjectured to support exotic quasiparticles, which may serve as basic ingredients for quantum technologies~\cite{nayak2008non}. However, 
the experimental demonstration is debated, and an unambiguous observation is still lacking. 
For example, current experimental evidence for the observation of Majorana fermions (MFs) is based on measurements of zero-bias peaks in the differential conductance which, however, may be attributable to other sources~\cite{pan2020physical}.
By contrast, thermodynamic quantities can unambiguously distinguish MFs from simpler excitations~\cite{cooper2009observable,viola2012thermoelectric,ben2013detecting}. In particular, the additional entropy due to a single Majorana fermion is $S = \frac{1}{2}k_{\rm{B}} \log 2$ -- 
 half that of a regular spin-degenerate state. This fractional entropy implies that information is stored non-locally across a pair of decoupled bound states.  
The measurement of a fractional entropy would therefore serve as a smoking-gun signature for exotic  quasiparticles~\cite{yang2009thermopower,sela2019detecting}.

In the context of the low-dimensional mesoscopic electronic systems predicted to host exotic quasiparticles, thermodynamic quantities are unfortunately difficult to measure experimentally. Techniques developed to measure extensive properties in bulk systems are inapplicable to identify small changes due to individual excitations over the large background phonon contributions. Thus observation of fractional entropy remains elusive.
Two indirect approaches to entropy measurement have been developed 
recently 
in nanoelectronic devices, although neither has as yet been applied to a system hosting exotic quasiparticles. One method utilizes thermopower measurements~\cite{kleeorin2019measure}, while the other exploits a Maxwell relation connecting entropy changes to \emph{charge} measurements~\cite{hartman2018direct,sela2019detecting,child2021entropy}.

In this Letter we discuss the latter approach in the context of charge-Kondo quantum dot devices~\cite{iftikhar2015two,iftikhar2018tunable}. These experimental systems are highly accurate circuit realizations of multichannel Kondo (MCK) models~\cite{iftikhar2015two,iftikhar2018tunable,nozieres1980p,affleck1991universal,matveev1991quantum,matveev1995coulomb,furusaki1995theory,mitchell2016universality}. Due to a frustration of Kondo screening at low temperatures, the two-channel charge-Kondo (2CK) model supports a Majorana fermion, with a residual `impurity' entropy $S_{\rm 2CK}=k_{\rm{B}}\log \sqrt{2}$, while the three-channel (3CK) model hosts a Fibonacci anyon as manifested by a residual entropy $S_{\rm 3CK}=k_{\rm B}\log \phi$, where $\phi=\tfrac{1}{2}(1+\sqrt{5})$ is the golden ratio~\cite{andrei1980diagonalization, vigman1980exact,affleck1991universal}. The central question we address in this work is: can these predicted MCK fractional entropies be measured and distinguished, taking into account the complexities and limitations of the experimental realization?

So far, only the electrical conductance of charge-Kondo devices has been measured experimentally~\cite{iftikhar2015two,iftikhar2018tunable,pouse2021exotic}, and so here we make use of this existing data by deriving novel exact relations between the conductance $G$ and the temperature-dependence of the dot occupation, $dN/dT$. Such relations are bijective and universal when the underlying physics is governed by a single energy scale, such as the Kondo temperature. The entropy is then extracted via a Maxwell relation \cite{hartman2018direct}. Assuming that the charge sensing protocol is minimally invasive, we conclude that the experimental observation of a nontrivial temperature scaling towards the ideal fractional values of the 2CK and 3CK entropy is entirely feasible. The method 
can be generalized to other systems, although the relations themselves are model-specific.


\emph{Multichannel Kondo models.--} 
A single spin-$\tfrac{1}{2}$ `impurity' coupled antiferromagnetically to one or more independent conduction electron channels represents an important paradigm in the theory of strongly correlated electrons.
At high temperatures, the impurity is effectively free, and so the impurity contribution to the total system entropy is $S_{\rm CB}=\log 2$ (setting $k_{\rm B} \equiv 1$ here and in the following).
The impurity becomes strongly entangled with conduction electrons in a surrounding `Kondo cloud' \cite{mitchell2011real,borzenets2020} at low temperatures $T\ll T_{\rm K}$, where $T_{\rm K}$ is 
the Kondo temperature. In the single-channel case, the conduction electrons exactly screen the impurity spin by formation of a many-body Kondo singlet, leaving zero residual entropy, $S_{\rm 1CK}=0$. For $k\ge 2$ channels, the frustration of Kondo screening results in an `overscreened' scenario ~\cite{andrei1980diagonalization, vigman1980exact,affleck1991universal} with a finite residual entropy $S_{k\rm{CK}} =\log\left\{2 \cos\left[ \pi/(2+k)\right]\right\}$, a hallmark of non-Fermi liquid (NFL) physics. 
In particular, the impurity in the 2CK model hosts an effective Majorana fermion at low temperatures \cite{emery1992mapping} with $S_{\rm 2CK}=\log\sqrt{2}$, while $S_{\rm 3CK}=\log \phi$, corresponding to a Fibonacci anyon, is predicted for 3CK~\cite{lopes2020anyons,komijani2020isolating}. 

\begin{figure*}[ht]
\includegraphics[width=1\textwidth]{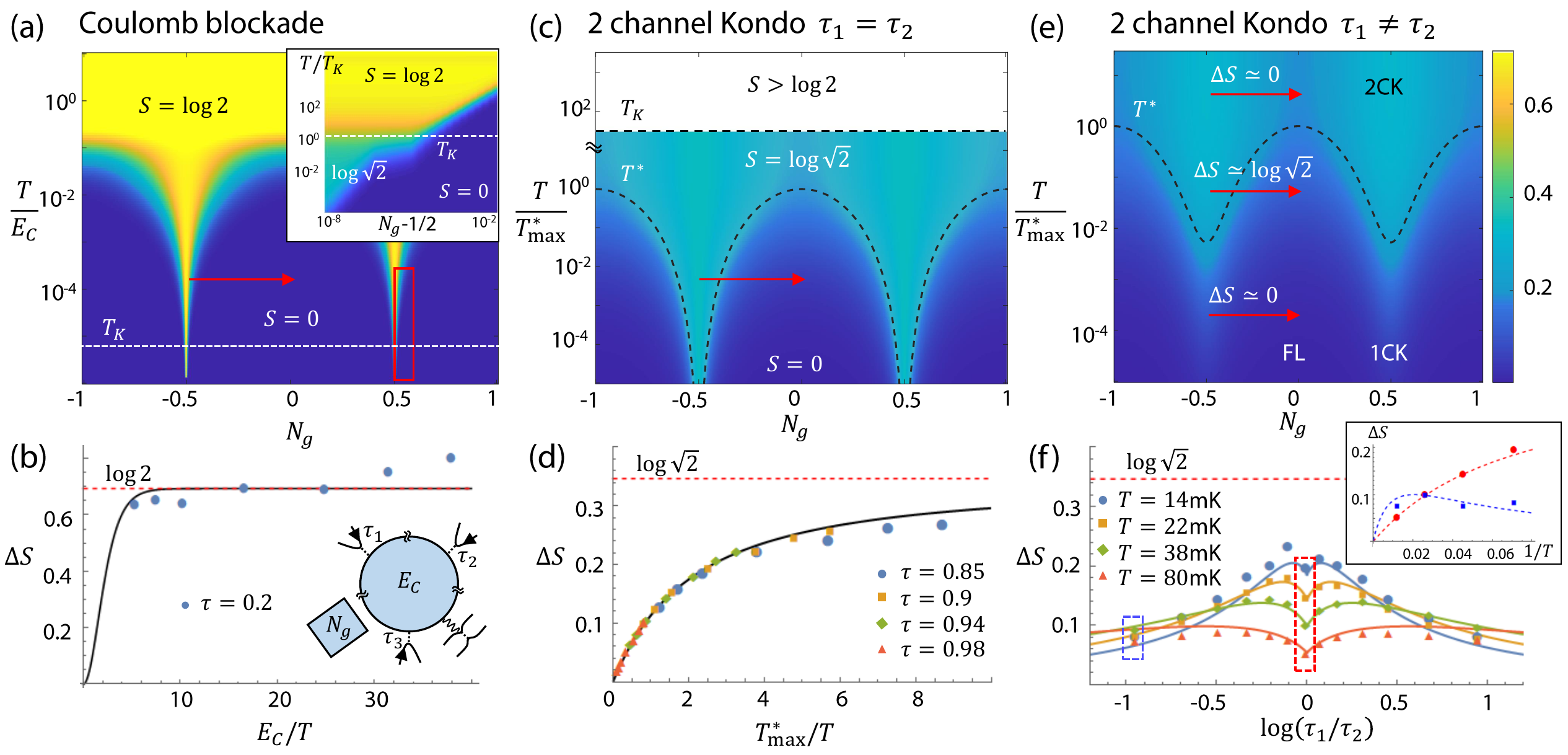}
\caption{
(a) Phase diagram of the charge $k$-channel Kondo device. For $T\gg T_K$ the entropy is $S=\log 2$ (yellow) at charge degeneracy $N_g=\pm\tfrac{1}{2}$, but $S=0$ (blue) in the Coulomb valley when $T\ll E_C$. Inset shows expanded view of red box, where the $k=2$ system maps to the 2CK model and the entropy for $T\ll T_K$ (obtained by NRG) approaches $\log \sqrt{2}$.
(b) Entropy difference along the red arrow in (a) obtained from experimental conductance data for $\tau=0.2$ and $T/$mK$=7.9, 9.5, 12, 18, 28.9, 40, 55$ via Eqs.~(\ref{eq:deltas},\ref{eq:dNdTCB}) (points) compared with Eq.~(\ref{eq:deltaSCB}) (line). Inset shows device schematic for $k=3$. 
(c) 2CK phase diagram and impurity entropy, showing the NFL-FL crossover on the scale of $T^*$ along the red arrow. 
(d) Entropy change extracted from the experimental 2CK conductance data via  Eqs.~(\ref{eq:deltas}), (\ref{eq:dNdT}) (points) as a function of $T^*_{\rm{M}}/T$ for different $\tau$, compared with Eq.~(\ref{eq:deltas2CK}) (line). 
(e) Phase diagram of channel-asymmetric 2CK system with $\tau_1\ne \tau_2$, showing  finite $T^*$ even at $N_g=\tfrac{1}{2}$. The low-$T$ suppression of impurity entropy results in the non-monotonic behavior of $\Delta S$ on decreasing $T$.
(f) Extracted entropy from experimental data as for (d) but for $\tau_1\ne \tau_2$, showing the NFL-FL crossover.
For $\log(\tau_1/\tau_2)<0$ ($>0$), we set $\tau_2$ ($\tau_1$) to $0.93$ and vary $\tau_1$ ($\tau_2$). Inset: temperature dependence of $\Delta S$  for $\tau_1=\tau_2$ (red) and $\tau_1\ne \tau_2$ (blue) corresponding to the dashed boxes.
}\label{fig:CB2CK}
\end{figure*}

\emph{Charge-Kondo realization.--}
The charge 2CK and 3CK effects were demonstrated~\cite{iftikhar2015two,iftikhar2018tunable} in a metallic quantum dot (QD) tunnel-coupled via quantum point contacts (QPCs) to two or three leads, as illustrated in the inset to Fig.~\ref{fig:CB2CK}(b). The device is operated in a large magnetic field such that the QD and leads are in the quantum Hall regime, and the QPCs each consist of a pair of counter-propagating spinless fermions, one incoming into the QD and the other outgoing from the QD.  
The Hamiltonian of the $k$-channel charge-Kondo model reads
\bea
H&=& \hbar v_\text{F}\sum_{j=1}^k \sum_{\nu=\text{in,out}} \int dx \psi^\dagger_{j\nu}i\partial_x \psi_{j\nu}+ E_C(\hat{N}-N_g)^2 \nonumber \\  &+&\hbar v_\text{F}\sum_{j=1}^k \left (r_j \psi_{j\text{in}}^\dagger(0)\psi_{j\text{out}}(0)+{\rm{H.c.}}\right ) .
\label{eq:Horigin}
\eea
The first term describes the free fermion modes $\psi_{j\nu}$ (with Fermi velocity $v_F$), while the second term describes the dot interactions with $E_C$ the charging energy and $\hat{N}$ the electron number  of the QD. $N_g$ is the gate voltage applied to the dot, normalized such that $N_g=1$ corresponds to addition of a single electron to the dot. 
The last term describes the reflection amplitude $r_j$ at each QPC, 
related to the  transmission coefficient $\tau_j$ as $1-\tau_j\simeq r_j^2$~\cite{chamon1997two,SM}. 

 The charge degeneracy at a Coulomb peak ($N_g=\tfrac{1}{2}$) maps to an effective impurity pseudospin-$\tfrac{1}{2}$~\cite{furusaki1995theory}; tunneling at the QPCs then correspond to pseudospin flip processes. The 2CK and 3CK models were found to accurately describe the experimental QD device with two and three leads by detailed comparison with electrical transport measurements \cite{iftikhar2015two,iftikhar2018tunable,furusaki1995theory,mitchell2016universality}. The thermoelectric response (as yet unmeasured in these systems) was predicted in Refs.~\cite{van2020wiedemann,*van2020electric,nguyen2020thermoelectric}.

 At the frustrated critical point, $r_j \equiv r$ (hence $\tau_j \equiv \tau$).  The Kondo temperature $T_{\rm K}$ is determined by the transmission; at a given temperature $T$, small $\tau$ implies $T\gg T_{\rm K}$ (where $T_K\propto E_C e^{-\pi^2/\sqrt{4\tau}}$) and so the system is in the classical Coulomb blockade (CB) regime, while for larger $\tau$ such that $T\ll T_{\rm K}$ (e.g. $T_K\propto E_C/(1-\tau)$ for 2CK~\cite{furusaki1995theory,iftikhar2015two}), the system exhibits the overscreened MCK effect.
 By detuning the gate voltage, a crossover is induced from the 
 NFL point with fractional entropy, to a 
 Fermi liquid (FL) state with quenched impurity entropy 
\cite{matveev1991quantum,matveev1995coulomb,furusaki1995theory,mitchell2016universality}. 

\emph{Maxwell Relation.--} 
We focus on the entropy change $\Delta S$ occurring as the gate voltage is swept from $N_g=0$ to $\tfrac{1}{2}$, corresponding to the crossover from the FL/trivial state with zero entropy, to the critical point which has fractional entropy for $T\ll T_{\rm K}$. From theory we therefore expect $\Delta S(T)$ to approach $S_{\rm 2CK}$ or $S_{\rm 3CK}$ for the two- and three-channel charge-Kondo devices as $T/T_{\rm K}$ is lowered. The Maxwell relation relates this change to the gate-voltage integral of $dN/dT$ viz,
\begin{align}
    \Delta S=2E_C \int_{0}^{1/2 }dN_g \frac{dN}{dT}.\label{eq:deltas}
\end{align}


\emph{Entropy and conductance-charge relation in CB.--} The system is in the CB regime for large QPC reflections, such that $E_C\gg  T\gg T_K$. The phase diagram in this regime is depicted in Fig.~\ref{fig:CB2CK}(a), where the red arrow denotes the line along which the entropy change is measured. Here we expect $\Delta S \simeq \log 2$ since $S_\text{imp}\simeq 0$ at the Coulomb valley, while for $N_g=\tfrac{1}{2}$ the impurity spin-$\tfrac{1}{2}$ remains largely unscreened. 

In the CB regime, both the conductance and the number of electrons in the dot can be obtained by classical rate equations,
\begin{equation}
\begin{split}
\label{eq:dNdTCB}
\frac{dN_\text{CB}}{dT}
=&\frac{1}{2T}\tanh\frac{E_C(N_g-\frac{1}{2})}{T}\frac{\frac{2E_C}{T}(N_g-\frac{1}{2})}{\sinh(\frac{2E_C}{T} (N_g-\frac{1}{2}))}\\
\equiv& \frac{1}{2T}\tanh\frac{E_C(N_g-\frac{1}{2})}{T}~\frac{G_\text{CB}}{G_\text{CB,max} },
\end{split}
\end{equation}
where $G_{\rm CB}$ is the CB conductance and $G_\text{CB,max}$ is the Coulomb peak conductance (itself a function of $\tau$ and $T$~\cite{iftikhar2015two,furusaki1995theory}). Physically, this relation follows from the fact that both observables are proportional to the dot density of states. In deriving Eq.~(\ref{eq:dNdTCB}) we assumed $T \ll E_C$ and retained only two dot charge states (therefore it is not periodic in $N_g$ and should be used only in the vicinity of a well-isolated CB peak). Within the same approximation, we can calculate the thermodynamic free energy  and hence obtain a prediction for $\Delta S$ directly,
\begin{align}
    \Delta S_\text{CB}=&\log 2 +\frac{E_C/T}{1+e^{-E_C/T}}-\log(1+e^{E_C/T}).\label{eq:deltaSCB}
\end{align}
This result also coincides with that obtained from Eq.~(\ref{eq:deltas}) using $dN_\text{CB}/dT$. Note that $\Delta S_\text{CB}\rightarrow\log 2$ as $T\rightarrow 0$, corresponding to the unscreened two-fold charge degeneracy of the dot in the CB regime.

We now utilize experimental data for the CB conductance \cite{iftikhar2015two}
and Eqs.~(\ref{eq:deltas}), (\ref{eq:dNdTCB}) to obtain an estimate for $\Delta S_\text{CB}$ in the experimental setting of the charge-Kondo device. The results are plotted in Fig.~\ref{fig:CB2CK}(b), comparing with Eq.~(\ref{eq:deltaSCB}) (line). The agreement is good at higher temperatures as expected, although experimental noise around the low-conductance signal in the CB regime naturally introduces 
errors. At lower temperatures, deviations due to Kondo renormalization are observed. To capture the behavior in the Kondo regime, one must use a more sophisticated theory to obtain the full conductance-charge relation. While the theoretical curve displays a decay of $\Delta S_\text{CB}$ to zero for temperature exceeding the charging energy, the present experimental data is restricted to temperatures up to $55$mK which is well below $E_C$. In the following we do this analytically for 2CK and numerically for 3CK.


\emph{Entropy and conductance-charge relation for 2CK.--} 
We consider now the 2CK case with $\tau_1 = \tau_2 \equiv \tau$ and $T\ll T_{\rm K}$ (large $\tau$ regime). A new energy scale is generated by gate-voltage detuning,  $T^*= T^*_{{\rm{M}}} \cos^2(\pi N_g)$, with $ T^*_{{\rm{M}}}= 8e^{\bf C} E_C (1-\tau)/\pi^2$ (and $\bf C$ the Euler constant). The NFL phase is stabilized for $T^*\ll T\ll T_K$, while for $T\ll T^*$  the system is in the zero-entropy, FL state -- see Fig.~\ref{fig:CB2CK}(c).  Thus, a fractional entropy change is expected along the red arrow, corresponding to the crossover between FL and NFL states. The model can be mapped to a resonant Majorana tunneling model~\cite{emery1992mapping,furusaki1995theory,SM}; this analytic solution allows us to obtain $dN/dT$ as well as the relation between $dN/dT$ and $G$ at large $\tau$: 
\begin{align}
    \frac{dN_\text{2CK}}{dT}&= \frac{T^*_\text{M}\sin(2\pi N_g)}{4E_C  T}\left (1-\frac{T^*}{2\pi T}\psi^{(1)}\left [ \frac{1}{2}+\frac{T^*}{2\pi T}\right] \right) \nonumber \\
&\equiv \frac{T^*_{\rm{M}}\sin(2\pi N_g)}{4E_C T} ~\frac{2G_\text{2CK}h}{e^2}.
\label{eq:dNdT}
\end{align}
For comparison, a direct 
calculation of the entropy yields 
\begin{equation}
\begin{split}
\Delta S_\text{2CK}=\frac{T^*_{\rm{M}}\left[\psi\left (\frac{1}{2} + \frac{T^*_{\rm{M}} }{2\pi T}\right )-1\right]}{2\pi T}-  \log\left [\frac{\Gamma\left (\frac{1}{2} +\frac{T^*_{\rm{M}}}{2\pi T}\right )}{\sqrt{\pi}}\right ]   ,\label{eq:deltas2CK}
\end{split}
\end{equation}
where $\psi(x)$ and $\psi^{(1)}(x)$ are the di-gamma and tri-gamma function, respectively.

We again use experimental conductance data \cite{iftikhar2015two} to obtain $dN/dT$ via Eq.~(\ref{eq:dNdT}), taking the experimental values of $E_C$ and $\tau$ -- but now in the nontrivial Kondo regime of the two-lead device. The entropy change is then extracted from Eq.~(\ref{eq:deltas}), and plotted in Fig.~\ref{fig:CB2CK}(d) for various values of $\tau$. 
The experimental data are seen to collapse accurately onto the predicted scaling form of Eq.~(\ref{eq:deltas2CK}) (solid line) for large $\tau$ where the above relation applies, with the entropy change tending to  $\Delta S \to S_{\rm 2CK}$ as the temperature is lowered. Appreciable deviations appear only for $\tau \lesssim 0.85$, for which the leading irrelevant operator must be taken into account in the theory~\cite{furusaki1995theory,affleck1991critical}. 

We note that the impurity entropy at $N_g=0$ does not reach zero even at the experimental base temperature of $T=7.9$mK for $E_C=300$mK, as reported in Ref.~\cite{iftikhar2015two}. Therefore the measured entropy change is about 80\%  of the ideal 2CK bound $S_{\rm 2CK}=\log\sqrt{2}$. Our results (in particular the temperature scaling) are consistent with the formation of a single Majorana fermion on the dot, but the bound would be better saturated for larger $E_C/T$.

Next we consider the channel-asymmetric case where $\tau_1\ne \tau_2$. On reducing the temperature, the system now flows from the  2CK critical point to a 1CK FL state with the more strongly coupled channel -- see Fig.~\ref{fig:CB2CK}(e). 
Extending the theory to this case, the crossover scale marked as a dashed line in Fig.~\ref{fig:CB2CK}(e) becomes \be
T^*=\frac{2}{\pi^2}e^{\bf C}E_C(2-\tau_1-\tau_2+2\sqrt{(1-\tau_1)(1-\tau_2)}\cos(4\pi N_g)) .
\ee
The relation between $dN/dT$ and the conductance Eq.~(\ref{eq:dNdT}) still holds once $(1-\tau)$ in the expression for $T^*_\text{M}$ is replaced by $\sqrt{(1-\tau_1)(1-\tau_2)}$.
This shows how the fractional entropy in the NFL phase is quenched to zero as the channel asymmetry drives the system to the 1CK FL regime.  
Applying this relation to the available experimental conductance data~\cite{iftikhar2015two} we obtain $\Delta S$ along this crossover. The entropy $\Delta S$ extracted in this way is in quantitative agreement with the direct 
calculation of the entropy, shown as solid lines in Fig.~\ref{fig:CB2CK}(f). Interestingly, $\Delta S$ is not a monotonic function of $\log(\tau_1/\tau_2)$. This is due to the fact that as one, e.g., decreases $\tau_2$, $T^*$ increases. As a consequence, the system is first driven closer to the 2CK fixed point, before turning over and flowing toward the 1CK fixed point. Similarly, $\Delta S$ is not a monotonic function of $T$, see Fig.~\ref{fig:CB2CK}(e) and inset to Fig.~\ref{fig:CB2CK}(f).


\begin{figure}[]
\includegraphics[width=1\columnwidth]{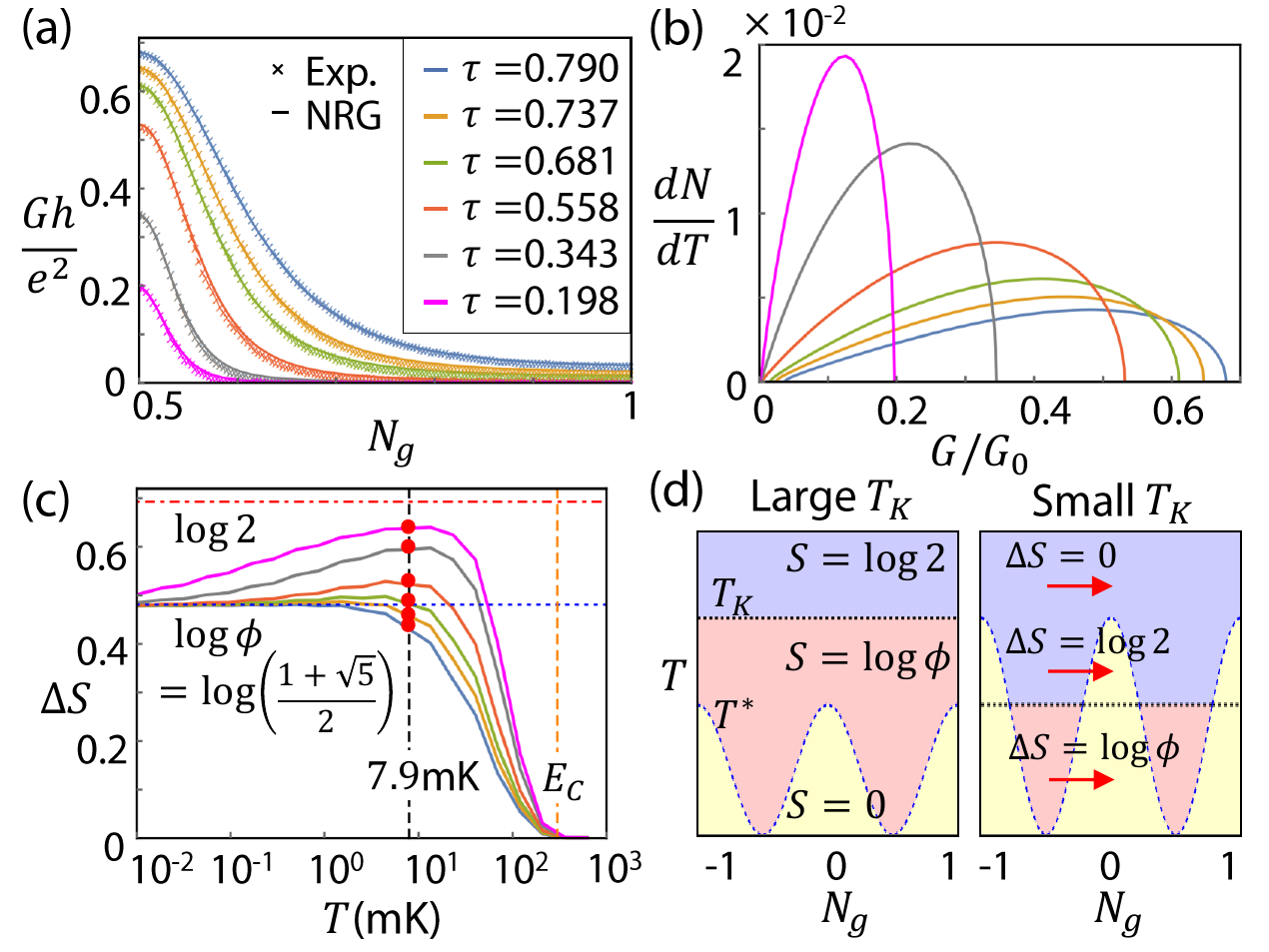}
\caption{(a)  Conductance $G$ for the 3CK system as a function of $N_g$ for different $\tau$ at $T=7.9$mK, comparing experimental data (points) with NRG calculations (lines). 
(b) Numerically-determined relation between $dN/dT$ and $G$ for the same model parameters as in (a). 
(c) 3CK entropy difference $\Delta S$ (points) from Eq.~(\ref{eq:deltas}) using $G$ from panel (a) and $dN/dT=f(G)$ from panel (b), compared with thermodynamic NRG results (lines). (d) Schematic phase diagrams for large transmission (left) and small transmission (right): CB regime in blue  ($T\gg T_K$, $S=\log 2$), FL in yellow ($T\ll T^*$, $S=0$), NFL in red ($T^*<T<T_K$, $S_{\rm 3CK}=\log\phi$).
} \label{fig:3CK}
\end{figure}

\emph{Entropy and conductance-charge relation for 3CK.--} 
We now turn to the critical 3CK system obtained by tuning $\tau_1=\tau_2=\tau_3 \equiv \tau$ in the three-lead charge-Kondo device \cite{iftikhar2018tunable}. Here the situation is more complex theoretically, as an analytical solution along the NFL to FL crossover is not known. Instead we employ a numerical solution using state-of-the-art numerical renormalization group (NRG) calculations \cite{wilson1975renormalization,*bulla2008numerical,weichselbaum2007sum,mitchell2014generalized,*stadler2016interleaved} to obtain the conductance $G$ as well as $dN/dT$ along the crossover. 

By comparison of NRG model predictions to conductance data~\cite{iftikhar2018tunable} we note that the experiment at large $\tau$ is not in the fully universal regime. In order to make our comparison with experiment quantitative, we generalize the standard 3CK model by retaining multiple dot charge states~\cite{SM}. For each experimental conductance curve we fit the model parameters to best match the conductance lineshapes. We consider a single (base) temperature of $T=7.9$mK, and transmissions ranging from $\tau=0.79$ to $\tau=0.198$ as in the experiment, which corresponds to the crossover regime from $T\ll T_K$ to $T\gg T_K$.

Figure~\ref{fig:3CK}(a) depicts the remarkable agreement between the experimental conductance curves and the NRG lineshapes over the entire range of $N_g$ for all values of the transmission considered. This again validates the theoretical model as an accurate description of the physical device. For these same parameters, NRG also yields $dN/dT$, and hence we deduce numerically the relation between $G$ and $dN/dT$, as shown in Fig.~\ref{fig:3CK}(b). 
Unlike the CB and 2CK cases (Eqs.~\eqref{eq:dNdTCB}, \eqref{eq:dNdT}), in 3CK, there is no simple relation between $dN/dT$ and $G$~\cite{SM}. Accordingly, for each set of parameters,  we use the numerically obtained  relation between $dN/dT$ and $G$  to determine the entropy difference $\Delta S$ via Eq.~(\ref{eq:deltas}). 
These are marked as the circle points in Fig.~\ref{fig:3CK}(c). 

The results are consistent with the entropy obtained from standard thermodynamic NRG calculations (Fig.~\ref{fig:3CK}(c), lines), which include the full temperature dependence. For larger transmission we find a direct crossover from $\Delta S=0$ to $\log \phi$ for $T\ll T_K$, corresponding to the 3CK NFL state hosting a Fibonacci anyon. 
Interestingly, for small $\tau$ we find that $\Delta S$ first approaches $\log 2$  for $T\gg T_K$ before approaching $\log \phi$ at the low-temperature limit.
 These two behaviours for large and small transmissions are illustrated in the two phase diagrams in  Fig.~\ref{fig:3CK}(d). For large $\tau$, due to large charge fluctuations, the 3CK Kondo temperature $T_K$ well exceeds the $N_g$ dependent crossover scale $T^*$. As a result, $\Delta S$ -- which probes gate-voltage sensitivity -- increases from $0$ to $\log \phi$ as $T$ is lowered below the crossover temperature $T^*$. For small transmission, however, there are effectively only two accessible charge states, as in the familiar spin Kondo problem. Then we have a large gate-voltage sensitivity, which acts as an effective magnetic field on the impurity pseudospin.  Similar to the situation shown in Fig.~\ref{fig:CB2CK}(a), above $T_{\rm K}$ the system is described by the CB theory, and below $T_K$ there is a FL--NFL crossover, leading to the non-monotonic behavior observed in Fig.~\ref{fig:3CK}(c) at small $\tau$.


\emph{Conclusion.--}
Nanoelectronic devices 
offer a controllable route to realizing anyonic quasiparticles. Their existence 
can be demonstrated through their fractional entropy. Here we develop an indirect route to the latter in 2CK and 3CK systems by (i) exploiting a Maxwell relation connecting the entropy change $\Delta S$ to the charge variation $dN/dT$, and (ii) deriving relations between $dN/dT$ and the measured electrical conductance $G$. Applying our methodology to existing conductance data for 2CK and 3CK charge-Kondo devices \cite{iftikhar2015two,iftikhar2018tunable}, we observe a scaling towards the expected nontrivial entropy value $S_{\rm 2CK}=\log\sqrt{2}$ for a Majorana anyon case. Likewise  our analysis of the 3CK model for various transmission values is consistent with $S_{\rm 3CK}=\log \phi$ for a Fibonacci anyon. Although experimental charge measurements on these systems are required for the incontrovertible demonstration of fractional quasiparticles, our analysis shows that existing experiments do operate in the necessary regime, and that the protocol for the observation of fractional entropy via the Maxwell relation is feasible. 
Entropy spectroscopy could serve as a smoking-gun probe of  exotic anyons in other more controversial systems such as Majorana wires~\cite{smirnov2015majorana,sela2019detecting} or other experimental systems realizing Kondo criticality \cite{potok2007observation,keller2015universal,mitchell2021so,*liberman2021so,pouse2021exotic,mebrahtu2013observation}.


\begin{acknowledgments}
\emph{Acknowledgments.--}
This  project  received  funding  from  European  Research Council (ERC) under the European Unions Horizon 2020 research and innovation programme under grant agreement No. 951541. AKM acknowledges funding from the Irish Research Council Laureate Awards 2017/2018 through Grant No. IRCLA/2017/169. AA and FP acknowledge support from the French RENATECH network and the French National Research Agency (ANR-16-CE30-0010-01 and ANR-18-CE47-0014-01). YM acknowledges support by the Israel  Science  Foundation  (grant  3523/2020). ES acknowledges support from ARO (W911NF-20-1-0013), the Israel Science Foundation grant number 154/19 and US-Israel Binational Science Foundation (Grant No. 2016255). 
\end{acknowledgments}


\bibliography{QIbasedNonAbelian}
\end{document}



\title{Fractional entropy of multichannel Kondo systems from conductance-charge relations:\\Supplemental Material}


\author{Cheolhee Han}
\affiliation{Raymond and Beverly Sackler School of Physics and Astronomy, Tel Aviv University, Tel Aviv 69978, Israel}
\author{Z. Iftikhar}
\affiliation{Universit\'e Paris-Saclay, CNRS, Centre de Nanosciences et de Nanotechnologies (C2N), 91120 Palaiseau, France}
\author{Yaakov Kleeorin}
\affiliation{Center for the Physics of Evolving Systems, 
University of Chicago , Chicago, IL, 60637, USA}
\author{A. Anthore}
\affiliation{Universit\'e Paris-Saclay, CNRS, Centre de Nanosciences et de Nanotechnologies (C2N), 91120 Palaiseau, France}
\affiliation{Universit\'{e} de Paris,  F-75006 Paris, France}
\author{F. Pierre}
\affiliation{Universit\'e Paris-Saclay, CNRS, Centre de Nanosciences et de Nanotechnologies (C2N), 91120 Palaiseau, France}
\author{Yigal Meir}
\affiliation{Department of Physics, Ben-Gurion University of the Negev, Beer-Sheva, 84105 Israel}
\author{Andrew K. Mitchell}
\affiliation{School of Physics, University College Dublin, Belfield, Dublin 4, Ireland}
\affiliation{Centre for Quantum Engineering, Science, and Technology, University College Dublin,  Dublin 4, Ireland}
\author{Eran Sela}
\affiliation{Raymond and Beverly Sackler School of Physics and Astronomy, Tel Aviv University, Tel Aviv 69978, Israel}

\maketitle


\section{Charge-2CK model: bosonization and refermionization}

We start with the original model for the charge Kondo system in Eq.~(1) of the main paper.
The electron number operator of the QD is defined as,
\be
\hat{N}=\sum_{j=1}^k \int_0^\infty dx[\psi^\dagger_{j\text{in}}(x)\psi_{j\text{in}}(x)+\psi^\dagger_{j\text{out}}(-x)\psi_{j\text{out}}(-x)].
\ee
The range corresponding to the QD is $0<x<\infty$ for the incoming modes and from $-\infty<x<0$ for the outgoing modes. 
Using bosonization, the fermion operators are represented by \cite{furusaki1995theory,le2002capacitance},
\begin{align}
\psi^\dagger_{j\nu}(x)=\sqrt{\frac{D}{\pi \hbar v_\text{F}}} \eta_{j\nu}e^{-i\phi_{j\nu}(x)}.
\end{align}
$D$ is bandwidth of the system and $\eta_{j\nu}$ is a local Majorana fermion, which satisfies the proper anticommutation relations  $\{\eta_{j\nu},\eta_{j'\nu'}\}=\delta_{j,j'}\delta_{\nu,\nu'}$ for the various species $(j=1,\cdots,k;\nu=\text{in, out})$ of fermions.  Equivalently there are four boson modes corresponding to the fermion modes, whose commutators are $[\phi_{j\nu}(x),\phi_{j'\nu'}(y)]=i\pi \text{sgn}(x-y)$ (we use a convention with right-movers only). 

We now specialize to the 2CK case $(k=2)$. The boson modes are transformed as
\be
\left(\begin{array}{l}
	\phi_C\\ \phi_I\\  \phi_1\\ \phi_2
\end{array}\right)=\frac{1}{2}
\left(\begin{array}{rrrr}
	1&-1&1&-1\\
	1&-1&-1&1\\
	1&1&1&1\\
	1&1&-1&-1
\end{array}\right)
\left(\begin{array}{c}
	\phi_{1\text{in}}\\ \phi_{1\text{out}}\\ \phi_{2\text{in}}\\ \phi_{2\text{out}}
\end{array}\right).
\ee
Also the Majorana fermions are transformed as
$\eta_C \eta_I=\eta_{1\text{in}} \eta_{1\text{out}}=\eta_{2\text{in}}\eta_{2\text{out}}$.
The Hamiltonian becomes
\begin{align}
H_0=&\sum_{i=1,2,C,I}\int dx\frac{\hbar v_\text{F}}{4\pi}(\partial_x\phi_i)^2,\nonumber\\
H_C=&\frac{E_C}{\pi^2} (\phi_C(0)+\pi N_g)^2,\\
H_B=&\frac{D}{\pi}\eta_C \eta_I  (r_1 e^{i\phi_C(0)}+r_2 e^{-i\phi_C(0)}) e^{i\phi_I(0)}+\text{H.C.}.\nonumber
\end{align}

If the charging energy $E_C$ exceeds all other energy scales, we can integrate out $\phi_C$~\cite{furusaki1995theory,le2002capacitance}. Subsequently,  we have $e^{i\phi_C(0)}\simeq \sqrt{\frac{2\gamma E_C }{D\pi}} e^{-i\pi N_g}$. We refermionize the model using $\psi^\dagger_I(0)=\sqrt{\frac{D}{\pi \hbar v_\text{F}}}\eta_I e^{-i\phi_I(0)} $. For convenience, we denote $r=\frac{r_1+r_2}{2}$ and $\bar{r}=\frac{r_1-r_2}{2}$. 
The Hamiltonian acquires a double-Majorana resonant level form
\begin{equation}
\begin{split}
    H=&\sum_{j=1,2}
    \frac{\hbar v_\text{F}}{4\pi}
    \int dx (\partial_x\phi_j)^2+\hbar v_\text{F}\int dx \psi_I^\dagger\partial_x\psi_I\nonumber\\ &+\sqrt{\frac{8\gamma E_C \hbar v_\text{F}}{\pi^2}}(r\cos(\pi N_g)\eta_C(\psi_I(0)-\psi_I^\dagger(0))\\ &+i\bar{r}\sin(\pi N_g)\eta_C(\psi_I(0)+\psi_I^\dagger(0)).\label{eq:HMethod}
\end{split}
\end{equation}
This simple form of the Hamiltonian of the non-trivial 2CK fixed point allows us to obtain analytic expressions for the thermodynamic entropy and for $dN/dT$.


\section{$dN/dT$ from conductance:\\CB and 2CK cases} 
\begin{figure}[t]
\includegraphics[width=0.95\columnwidth]{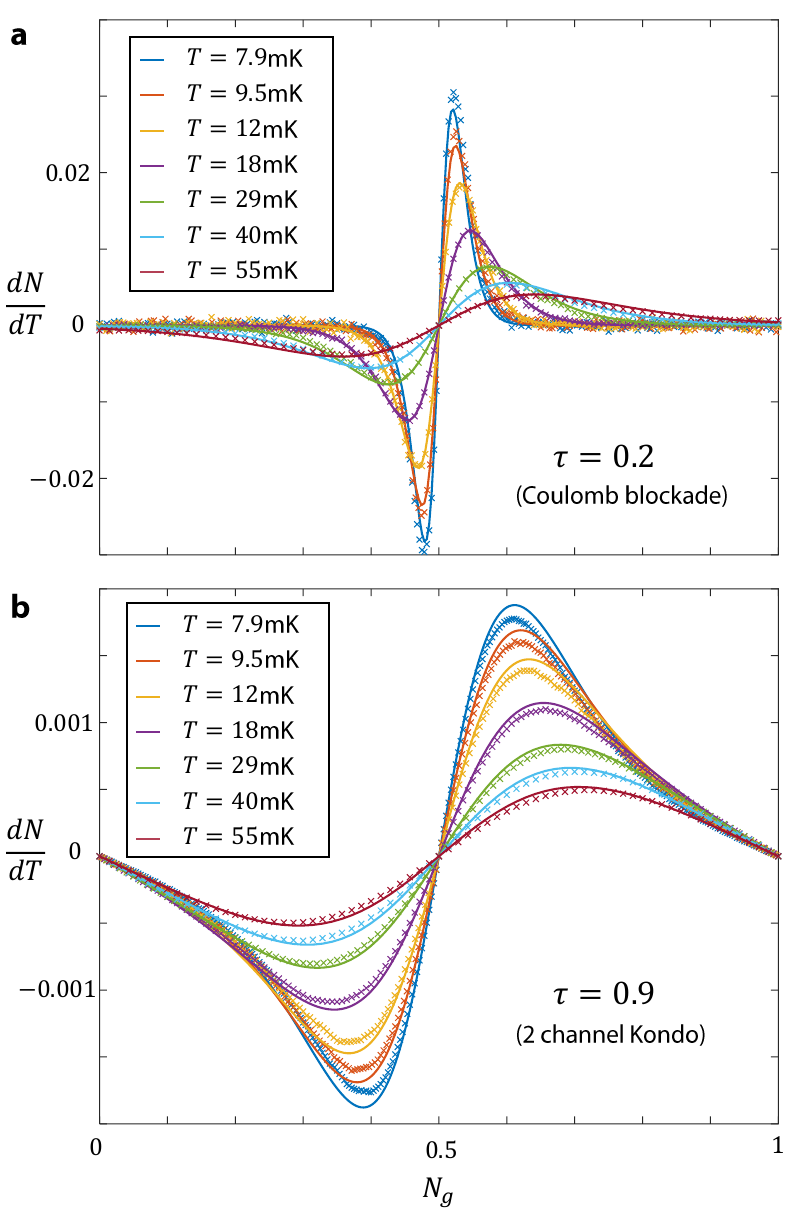}
\caption{(a) $dN/dT$ for Coulomb blockade regime ($\tau=0.2$). Solid lines are theoretical curves, and points show the experimentally extracted $dN/dT$ from Eq.~(3). (b) $dN/dT$ for 2 channel Kondo regime ($\tau=0.9$). Solid lines are theoretical curves, and points show the experimentally extracted $dN/dT$ from Eq.~(5). 
} \label{fig:dndtCB2CK_supp}
\end{figure}

In Fig.~\ref{fig:dndtCB2CK_supp} we compare $dN/dT$ obtained from experimental conductance data either in the CB or 2CK regimes, using the relations in Eqs.~(3) or (5) of the main paper, with the theoretically derived results. In the CB regime, following Ref.~\cite{glazman2003coulomb}, we use,
\begin{equation}
    \frac{dN}{dT}=\frac{1}{2T}\tanh\left[\frac{E_C(N_g-1/2)}{T}\right]\frac{2E_C(N_g-1/2)}{T\sinh(2E_C (N_g-1/2)/T)}.\label{eq:dNdTCBexpress}
\end{equation}
For the 2CK regime, we use 
\begin{widetext}\begin{equation}
\frac{dN}{dT}=\frac{2\gamma(r^2-\bar{r}^2)}{\pi^2 T}\sin(2\pi N_g)\left(1-\frac{\frac{8\gamma E_C}{\pi^2}[ r^2\cos^2(\pi N_g)+ \bar{r}^2\sin^2 (\pi N_g)]}{2\pi T}\psi^{(1)}(\frac{1}{2}+\frac{\frac{8\gamma E_C}{\pi^2}[ r^2\cos^2(\pi N_g)+\bar{r}^2\sin^2 (\pi N_g)]}{2\pi T})\right),\label{eq:dNdT2CKexpress}
\end{equation}
\end{widetext}
whose derivation is given below. For each QPC, the transmission $0 \le \tau_i \le 1$ $(i=1,2,3)$ is related to the reflection amplitude $r_i$ according to~\cite{chamon1997two} $1-\tau_i=\frac{r_i^2}{(1+r_i^2/4)^2}$.

Experimentally extracted values fit with the theory quantitatively well. $\Delta S$ shown in Fig.~1(b) and (c) of the main paper is obtained by
\begin{equation}
    \Delta S=\frac{1}{2}\Big[-\int_0^{1/2}dN_g \frac{dN}{dT}+\int_{1/2}^{1}dN_g \frac{dN}{dT}\Big],
\end{equation}
or averaging over two entropy differences, $S(N_g=1/2)-S(N_g=0)$ and $S(N_g=1/2)-S(N_g=1)$ to reduce experimental error.

\section{Derivation of Eq.~(5)}
Here we calculate the expectation value of the number of electrons in the quantum dot for the 2CK model.
We start by writing the Hamiltonian Eq.~(1) in a simpler form. We first rotate the fermion modes as
\begin{equation}
\begin{split}
    \chi_{I}=&\frac{i}{\sqrt{2}}\frac{1-i\frac{\bar{r}}{r}\tan(\pi N_g)}{\sqrt{1+\frac{\bar{r}^2}{r^2}\tan^2(\pi N_g)}}\psi_{I}^\dagger-\frac{i}{\sqrt{2}}\frac{1+i\frac{\bar{r}}{r}\tan(\pi N_g)} {\sqrt{1+\frac{\bar{r}^2}{r^2}\tan^2(\pi N_g)}} \psi_{I},\\
    \bar{\chi}_{I}=&\frac{1}{\sqrt{2}}\frac{1-i\frac{\bar{r}}{r}\tan(\pi N_g)}{\sqrt{1+\frac{\bar{r}^2}{r^2}\tan^2(\pi N_g)}}\psi_{I}^\dagger+\frac{1}{\sqrt{2}}\frac{1+i\frac{\bar{r}}{r}\tan(\pi N_g)}{\sqrt{1+\frac{\bar{r}^2}{r^2}\tan^2(\pi N_g)}}\psi_{I}.
\end{split}\label{eq:fermion_trans1}
\end{equation}
Using this notation, the Hamiltonian becomes
\begin{align}
H=&\sum_{\nu=1,2}\frac{\hbar v_\text{F}}{4\pi}\int dx (\partial_x\phi_{\nu})^2 \nonumber\\ & +i\hbar v_\text{F}\int dx \bar{\chi}_I \partial_x \bar{\chi}_I+i\hbar v_\text{F}\int dx \chi_I\partial_x \chi_I \nonumber\\
&+i4\sqrt{\frac{\gamma E_C \hbar v_F}{\pi^2}\big[r^2\cos^2(\pi N_g)+\bar{r}^2\sin^2(\pi N_g)\big]}\eta_C \chi_I(0).\label{eq:Hmapped}
\end{align}
From this model  we can obtain the thermodynamic entropy as~\cite{sela2019detecting,rozhkov1998impurity},
\begin{equation}
\begin{split}
    S=\int_{-E_C}^{E_C} \frac{d\omega}{2\pi}&\frac{\frac{8\gamma E_C\hbar v_F}{\pi^2}(r^2\cos^2(\pi N_g)+\bar{r}^2\sin^2(\pi N_g))}{\omega^2+\big[\frac{8\gamma E_C\hbar v_F}{\pi^2} (r^2\cos^2(\pi N_g)+\bar{r}^2\sin^2(\pi N_g))\big]^2} \\ & \times \Big[\frac{\omega/T}{1+e^{\omega/T}}+\log(1+e^{-\omega/T})\Big].\label{eq:thermoentropy}
\end{split}
\end{equation}
Equation~\eqref{eq:thermoentropy} is used for Figs.~1(c) and (e). For Fig.~1(c) we use $r=0.3987$, and for Fig.~1(e) we use $r=0.3962$ and $\bar{r}=0.1269$. 

The average electron number of the dot can be obtained using the relation
\begin{align}
N=N_g-\frac{1}{2E_C}\frac{\partial F}{\partial N_g}.
\label{eq:numberdef}
\end{align}

After computing  $\partial_{N_g}F$, the charge deviation becomes,
\begin{widetext}\begin{align}
     N=N_g- \sqrt{\frac{2\gamma\hbar v_\text{F}}{E_C}}\int_{-E_C}^{E_C}\frac{d\omega}{2\pi} \Big[-r\sin(\pi N_g)\langle\eta_C(\psi_I(0)-\psi^\dagger_I(0))\rangle(\omega) +i\bar{r}\cos(\pi N_g)\langle\eta_C(\psi_I(0)+\psi^\dagger_I(0))\rangle(\omega)\Big].
\end{align}
The energy integration is bounded by $E_C$, which is a high energy cutoff appearing after integrating out $\phi_C$. 
We can compute the correlators  $\langle \eta_C\psi_I(0)\rangle$ or $\langle \eta_C\psi^\dagger_I(0)\rangle$ by changing the basis back as in Eq.~(\ref{eq:fermion_trans1}).

Combining  Eqs.~\eqref{eq:numberdef} and \eqref{eq:fermion_trans1}, the number expectation value becomes
\begin{align}
N=&N_g+2 (r^2-\bar{r}^2)\sqrt{\frac{\gamma\hbar v_\text{F}}{E_C}}\sin(\pi N_g)\cos(\pi N_g)\frac{1}{\sqrt{r^2\cos^2(\pi N_g)+\bar{r}^2\sin^2(\pi N_g)}} \int_{-E_C}^{E_C}\frac{d\omega}{2\pi}(-i)\langle \eta_C\chi_I(0)\rangle.
\end{align}
Then $dN/dT$ becomes
\begin{align*}
\frac{dN}{dT}=&2 (r^2-\bar{r}^2)\sqrt{\frac{\gamma\hbar v_\text{F}}{E_C}}\sin(\pi N_g)\cos(\pi N_g)\frac{1}{\sqrt{r^2\cos^2(\pi N_g)+\bar{r}^2\sin^2(\pi N_g)}} \partial_T\int\frac{d\omega}{2\pi}(-i)\langle \eta_C\chi_I(0)\rangle.\label{eq:dNdT_asym}
\end{align*}

Next we calculate the conductance. The current operator is
\begin{align}
I=&e\partial_t\frac{1}{2}\Big(\int_{-\infty}^0 dx( \psi^\dagger_{1\text{in}}\psi_{1\text{in}}-\psi^\dagger_{2\text{in}}\psi_{2\text{in}})+\int_0^\infty dx( \psi^\dagger_{1\text{out}}\psi_{1\text{out}}-\psi^\dagger_{2\text{out}}\psi_{2\text{out}})\Big)\nonumber\\
=&\frac{e}{2\pi}\partial_t\phi_I(0)=e\bar{\chi}_I(0)\chi_I(0).
\end{align}
The conductance is 
\begin{align}
    G=&\frac{i}{2\hbar}\int dt \,t\langle [I(t),I(0)]\rangle=-\frac{e^2}{2\hbar}\int \frac{d\omega}{2\pi}A(\chi_I(0),\chi_I(0);\omega)A(\bar{\chi}_I(0),\bar{\chi}_I(0);-\omega)\partial_\omega n_F(\omega/T)\nonumber\\
    =&-\frac{e^2}{2\hbar}\int \frac{d\omega}{2\pi}A(\chi_I(0),\chi_I(0);\omega)\partial_\omega n_F(\omega/T)=\frac{e^2}{2\hbar}\int \frac{d\omega}{2\pi}A(\chi_I(0),\chi_I(0);\omega)\frac{T}{\omega}\partial_T n_F(\omega/T)\nonumber\\
=&\frac{e^2}{2\hbar}T\partial_T \int \frac{d\omega}{2\pi}\frac{1}{\omega}\langle \chi_I(0)\chi_I(0)\rangle(\omega)\nonumber\\
=&\frac{e^2}{2\hbar}T\frac{1}{\sqrt{\frac{16\gamma E_C }{\hbar v_F \pi^2}\big[r^2\cos^2(\pi N_g)+\bar{r}^2\sin^2(\pi N_g)\big]}}\partial_T\int \frac{d\omega}{2\pi}(-i)\langle \eta_C\chi_I(0)\rangle(\omega).\label{eq:cond_asym}
\end{align} 
Here $A(\chi_I(0),\chi_I(0);\omega)$ is spectral function of operator $\chi_I(0)$ at energy $\omega$. In the calculation, we use $A(\bar{\chi}_I(0),\bar{\chi}_I(0);
\omega)=1$ for non-interacting $\bar{\chi}_I$, and the equation of motion of $\chi_I$,
\begin{align}
    \sqrt{\frac{16\gamma E_C }{\hbar v_F \pi^2}\big[r^2\cos^2(\pi N_g)+\bar{r}^2\sin^2(\pi N_g)\big]}\chi_I(0,\omega)=-i\omega \eta_C(\omega).
\end{align}
Then
\begin{align}
    \frac{dN}{dT}=\frac{2\gamma}{\pi^2 T}(r^2-\bar{r}^2)\sin(2\pi N_g)\frac{2G}{G_0},
\end{align}
where again $G_0=e^2/h$.

Now we obtain the expressions for $dN/dT$ and $G$. From the correlators
\begin{equation}
    \begin{split}
        \langle \eta_C\chi_{I}(0)\rangle(\omega)&=\frac{i\omega\sqrt{\frac{16\gamma E_C }{\hbar v_F \pi^2}\big[r^2\cos^2(\pi N_g)+\bar{r}^2\sin^2(\pi N_g)\big]}}{\omega^2+\Big[\frac{8\gamma E_C }{\pi^2}\big[r^2\cos^2(\pi N_g)+\bar{r}^2\sin^2(\pi N_g)\big]\Big]^2}\frac{1}{1+e^{-\omega/T}},\\
    \langle \eta_C\bar{\chi}_{I}(0)\rangle(\omega)&=0,
    \end{split}\label{eq:correlator2CK}
\end{equation}
we can integrate Eq.~\eqref{eq:dNdT_asym} and Eq.~\eqref{eq:cond_asym}, leading to
\begin{align}
\frac{dN}{dT}=\frac{2\gamma(r^2-\bar{r}^2)}{\pi^2 T}\sin(2\pi N_g)\Big(1-\frac{\frac{8\gamma E_C}{\pi^2}( r^2\cos^2(\pi N_g)+ \bar{r}^2\sin^2 (\pi N_g))}{2\pi T}\psi^{(1)}(\frac{1}{2}+\frac{\frac{8\gamma E_C}{\pi^2}( r^2\cos^2(\pi N_g)+\bar{r}^2\sin^2 (\pi N_g))}{2\pi T})\Big),\label{eq:dNdTexpress}
\end{align}
and
\begin{align}
G=\frac{G_0}{2}\Big(1-\frac{\frac{8\gamma E_C}{\pi^2}( r^2\cos^2(\pi N_g)+ \bar{r}^2\sin^2 (\pi N_g))}{2\pi T}\psi^{(1)}(\frac{1}{2}+\frac{8\frac{\gamma E_C}{\pi^2}( r^2\cos^2(\pi N_g)+\bar{r}^2\sin^2 (\pi N_g))}{2\pi T})\Big).\label{eq:condexpress}
\end{align}
Equation~(5) of the main text relating $dN/dT$ with $G$ follows from Eqs.~(\ref{eq:dNdTexpress}) and (\ref{eq:condexpress}).

The extracted entropy $\Delta S$ is obtained as
\begin{align}
\Delta S=-\frac{4\gamma E_C}{\pi^3 T}(r^2-\bar{r}^2)+\log\Big[\frac{\Gamma(\frac{1}{2}+\frac{4\gamma E_C \bar{r}^2}{\pi^3 T})}{\Gamma(\frac{1}{2}+\frac{4\gamma E_C r^2}{\pi^3 T})}\Big]+\frac{4\gamma E_C}{\pi^3 T}\Big[r^2 \psi^{(0)}(\frac{1}{2}+\frac{4\gamma E_C r^2}{\pi^3 T})-\bar{r}^2 \psi^{(0)}(\frac{1}{2}+\frac{4\gamma E_C \bar{r}^2}{\pi^3 T})\Big],
\end{align}
which is used for the solid lines in Fig.~1(d) and (f).
\end{widetext}


\section{Leading irrelevant correction to Eq.~(5)}
Here we calculate the leading irrelevant corrections to $dN/dT$ and $G$. The lowest order corrections are
\begin{align}
\label{deltadndt}
\delta \left( \frac{dN}{dT} \right)=\frac{2r^2\gamma}{\pi^2}\sin(2\pi N_g)\frac{\pi^4 T}{4E_C^2}\Big[\frac{1}{12}+2\log\Big(\frac{2E_C}{\pi T}\Big)\Big],
\end{align}
\begin{align}
\label{deltaG}
\delta G=-\frac{G_0}{2}\frac{\pi^3\gamma T}{16E_C}r^2\sin^2(\pi N_g).
\end{align}
Equation (\ref{deltaG}) is derived in Ref.~\cite{furusaki1995theory}. Below we will derive Eq.~(\ref{deltadndt}). By inspection, we see that these corrections violate the relation in Eq.~(5). Therefore Eq.~(5) holds only for small $r$, small $T$, and large $E_C$.

 To compute the leading irrelevant correction to $dN/dT$, we use the bosonized description and treat the reflection as a perturbation. The leading irrelevant contribution to the conductance is computed within this framework in Ref.~\cite{furusaki1995theory}. 
 
The Lagrangian of the system is $\mathcal{L}=\mathcal{L}_0+\mathcal{L}_C+\mathcal{L}_B$ where
\begin{align}
\mathcal{L}_0=&\sum_{i=1,2,I,C}\frac{\hbar v_\text{F}}{4\pi}\int dx \partial_x\phi_i( \partial_t\phi_i-\partial_x\phi_i),\nonumber\\
\mathcal{L}_C=&-\frac{E_C}{\pi^2}(\phi_C(0)+\pi N_g)^2,\\
\mathcal{L}_B=&-\frac{Dr}{\pi} e^{i(\phi_I(0)+\phi_C(0))}+e^{i(\phi_I(0)-\phi_C(0))}+\text{H.C.}.\nonumber
\end{align}
We  Fourier transform the boson fields as
\begin{align}
\phi_i(x,\tau)=\sqrt{\frac{T}{L}}\sum_{\omega_n,q}\phi_i(q,\omega_n) e^{iqx-i\omega_n\tau},
\end{align}
where $\omega_n=2\pi T n/\hbar$, $L$ is system size, and $\tau$ is imaginary time.
The bare action is transformed to
\begin{align}
\mathcal{S}_0=-\sum_{\omega_n,q}-\frac{q(q-i\omega_n)}{4\pi}\phi_C(q,\omega_n)\phi_C(-q,-\omega_n).
\end{align}
We now focus on the $\phi_C$ and $\phi_I$ fields. Heading towards a perturbative expansion in the backscattering term $\mathcal{L}_B$, we integrate these two fields away from $x=0$. Neglecting $\mathcal{L}_B$ at first, and noticing that the fields $\phi_{1,2}$ are decoupled, we obtain
\begin{align}
\mathcal{Z}_{\phi_I}=&\int\mathcal{D}\phi_I \exp\bigg[-\sum_{\omega_n}\frac{|\omega_n|}{2\pi}\phi_I(\omega_n)\phi_I(-\omega_n)\bigg],\nonumber\\
\mathcal{Z}_{\phi_C}=&\int\mathcal{D}\phi_C \exp\bigg[-\sum_{\omega_n}\Big(\Big[\frac{|\omega_n|}{2\pi}+\frac{E_C}{\pi^2}\Big]\phi_C(\omega_n)\phi_C(-\omega_n)\Big) \nonumber \\ &+\frac{2E_C}{\pi}\frac{1}{\sqrt{T}}N_g\phi_C(\omega_n=0)-\frac{E_C}{T} N_g^2\bigg].
\end{align}
Now we expand the perturbation $\mathcal{L}_B$. Then the partition function can be expanded as $\mathcal{Z}\simeq \mathcal{Z}_1+\frac{D^2 r^2}{2\pi^2} \mathcal{Z}_{2}$, where $\mathcal{Z}_1=\mathcal{Z}_{\phi_I}\mathcal{Z}_{\phi_C}$, and $\mathcal{Z}_2$ is written as
\begin{widetext}
\begin{align}
\mathcal{Z}_2=&\sum_{\xi_1\xi_2=\pm1}\int_0^{\frac{\hbar}{T}} d\tau_1 d\tau_2\int \mathcal{D}\phi_I\mathcal{D}\phi_C\exp\bigg[-\sum_{\omega_n}\Big(\frac{|\omega_n|}{2\pi}\phi_I(\omega_n)\phi_I(-\omega_n)-i\sqrt{T}\xi_1(e^{-i\omega_n\tau_1}- e^{-i\omega_n\tau_2})\phi_I(\omega_n)\Big)\bigg]\nonumber\\
\times&\exp\bigg[-\sum_{\omega_n}\Big([\frac{|\omega_n|}{2\pi}+\frac{E_C}{\pi^2}]\phi_C(\omega_n)\phi_C(-\omega_n)-i\sqrt{\frac{T}{\hbar}}\xi_1(e^{-i\omega_n\tau_1}-\xi_2 e^{-i\omega_n\tau_2})\phi_C(\omega_n)\Big)+\frac{2E_C N_g}{\pi\sqrt{T/\hbar}}\phi_C(\omega_n=0)- \frac{E_C}{T} N_g^2\bigg].\nonumber
\end{align}
Here $\xi_1$, $\xi_2=\pm 1$ indicate the tunneling directions ($\xi_1$) and tunneling positions ($\xi_2$).
From Eq.~\eqref{eq:numberdef}, the electron's number in the quantum dot is
\begin{align}
N=&N_g-\frac{T}{2E_C}\frac{1}{\mathcal{Z}}\frac{d \mathcal{Z}}{dN_g}\simeq N_g- \frac{2r^2\gamma}{\pi^2}\sin(2\pi N_g)   \frac{T}{\hbar}\int_0^{\frac{\hbar}{T}} d\tau_1\int_0^{\frac{\hbar}{T}} d\tau_2 \frac{\pi T}{\sin(\frac{\pi T}{\hbar}|\tau_1-\tau_2|)}\bigg[1-\frac{\pi^4 T^2}{4E_C^2\sin^2(\frac{\pi T}{\hbar}(\tau_1-\tau_2))}\bigg].\label{eq:number_pert}
\end{align}
To integrate  Eq.~\eqref{eq:number_pert}, we introduce the infinitesimal time $\tau_0$, which corresponds to the inverse of the bandwidth (or $\hbar/E_C$),
\begin{align}
    \int_{0}^{\hbar/T}d\tau \frac{\pi T}{\sin(\pi T\tau/\hbar)}=&\int_{\tau_0}^{\hbar/T-\tau_0}d\tau \frac{\pi T}{\sin(\pi T\tau)/\hbar}=2 \hbar \log(\frac{2}{\tau_0\pi T}),\\
    \int_{0}^{\hbar/T}d\tau \frac{(\pi T)^3}{\sin(\pi T\tau/\hbar)^3}=&\int_{\tau_0}^{\hbar/T-\tau_0}d\tau \frac{(\pi T)^3}{\sin(\pi T\tau)/\hbar}=\hbar^3\Big[\tau_0^2T^2 -\frac{\pi^2}{6} + \pi^2\log(\frac{2\hbar}{\pi\tau_0 T})\Big].
\end{align}
After integrating $\tau_1$ and $\tau_2$ we have
\begin{align}
N\simeq N_g-2\sin(2\pi N_g) \frac{r^2}{\pi^2}\gamma \log(\frac{2E_C}{\pi T})-2\sin(2\pi N_g) \gamma r^2[\frac{1}{8}-\frac{\pi^2T^2}{6E_C^2} +\frac{\pi^2T^2}{4E_C^2}\log(\frac{2E_C}{\pi T})],
\end{align}
and $dN/dT$ becomes
\begin{align}
\frac{dN}{dT}\simeq\frac{2r^2\gamma}{\pi^2}\sin(2\pi N_g)\frac{1}{T}+\frac{2r^2\gamma}{\pi^2}\sin(2\pi N_g)\frac{\pi^4 T}{4E_C^2}[\frac{1}{12}+2\log(\frac{2E_C}{\pi T})))].
\end{align}
The first term corresponds to a Taylor expansion for small $r$ of our result in Eq.~\eqref{eq:dNdTexpress} (with $\bar{r}=0$). 
The second part corresponds to the leading irrelevant corrections in Eq.~(\ref{deltadndt}).
\end{widetext}


\section{Details of NRG calculations\\for generalized 3CK model}
To simulate the experimental three-lead charge-Kondo system using NRG, we implement a generalization of the standard 3CK model to include multiple dot charging states and a finite charging energy $E_C$. Following Matveev \cite{furusaki1995theory} we identify three distinct channels of conduction electrons $\alpha=1,2,3$ around each of the three QPCs connecting the dot to the leads. We label electrons on the physical leads as $\sigma=\uparrow$ and electrons on the dot as $\sigma=\downarrow$ (even though in reality the electrons are effectively spinless due to the large applied magnetic field). Without interactions, the three channels are to a good approximation independent because of the decohering Ohmic contact \cite{iftikhar2018tunable}, which gives a long dwell time for electrons on the dot. However, tunneling events at the QPCs become correlated due to the dot charging energy $E_C$. Tunneling of electrons onto or off the dot changes the number of dot electrons $N$, and hence its electrostatic energy due to the Coulomb interaction. A gate voltage $V_{g}$ is used to tune the average dot filling, and hence $N_g$.

The Hamiltonian reads, $H_{\rm 3CK} = \sum_{\alpha}( H_{\rm leads}^{\alpha} + H_{\rm dot-lead}^{\alpha} ) + H_{\rm int} + H_{\rm gate}$. Here, 
\begin{equation}\label{eq:H3CK_leads}
H_{\rm leads}^{\alpha}=\sum_{k,\sigma}\epsilon_k^{\phantom{\dagger}} c_{\alpha k \sigma}^{\dagger}c_{\alpha k \sigma}^{\phantom{\dagger}} \;,
\end{equation}
describes each effective spinfull conduction electron channel. For simplicity, we take equivalent leads with a constant density of states $\nu=1/2D$ inside a band of half-width $D$. 

Electronic tunneling between the dot and lead $\alpha$ is described by,
\begin{equation}\label{eq:H3CK_dl}
H_{\rm dot-lead}^{\alpha}=J_{\alpha}^{\phantom{\dagger}} (c_{\alpha \downarrow}^{\dagger}c_{\alpha \uparrow}^{\phantom{\dagger}}\hat{S}^+ + c_{\alpha \uparrow}^{\dagger}c_{\alpha \downarrow}^{\phantom{\dagger}}\hat{S}^-)  \;,
\end{equation}
where $c_{\alpha\sigma}=\sum_k a_k c_{\alpha k \sigma}$ is a localized lead orbital at the dot position. We define $\hat{S}^+ = \sum_{N_d=N_0-\bar{N}}^{N_0+\bar{N}-1} |N_d+1\rangle\langle N_d |$ as an operator that increases the number of dot electrons by one unit (and correspondingly $\hat{S}^- = (\hat{S}^+)^{\dagger}$ decreases the dot charge), which thereby keeps track of the electron localization in either lead or dot around each QPC. Here $N_0$ is some fixed reference number of electrons, while $\bar{N}$ determines the number of accessible dot charge states. Formally $N_0,\bar{N}\rightarrow \infty$, however in practice a finite number of charge states $\bar{N}$ can be retained in the NRG calculations, provided the QPC transmission is not too high, and the temperature is low enough compared with the charging energy $E_C$. One can check \emph{post hoc} that the results of NRG calculations are converged with respect to increasing $\bar{N}$ for a given set of physical model parameters.

Electron interactions on the dot are embodied by the charging term,
\begin{equation}\label{eq:H3CK_int}
H_{\rm int}=E_C (\hat{N}_d - N_0)^2 \;,
\end{equation}
with $\hat{N}_d=\sum_{\alpha,k}c_{\alpha k \downarrow}^{\dagger}c_{\alpha k \downarrow}^{\phantom{\dagger}} $ the total dot number operator such that $\hat{N}_d|N_d\rangle = N_d|N_d\rangle$. With the gate voltage term $H_{\rm gate}=-eV_g \hat{N}_d$, we may write $H_{\rm int}+H_{\rm gate}=E_C(\hat{N}_d-N_0-N_g)^2$ up to an irrelevant overall constant, with $eV_g=2E_C N_g$. In the following we refer only to $N_g$, but it should be understood that this is tunable in practice through $V_g$. 

Note that for half-integer $N_0$ and $E_C\to \infty$, only the lowest two dot charge states survive, and the model becomes equivalent to a spin-anisotropic version of the standard spin-3CK model. However, in this work, we use a finite value of $E_C$ and retain multiple charge states. Specifically, we take $E_C=300$mK and $T=7.9$mK as in the experiment, and set $D/E_C=10$. We choose $N_0$ to be half-integer and take $\bar{N}=5$, meaning that in total 10 dot charge states are retained in the NRG calculations. We have confirmed that the results are insensitive to further increasing $D/E_C$ and $\bar{N}$. We focus on the properties at the critical point, for which we have equal couplings, $J_{\alpha}\equiv J$. 

\begin{figure}[t]
\includegraphics[width=1\columnwidth]{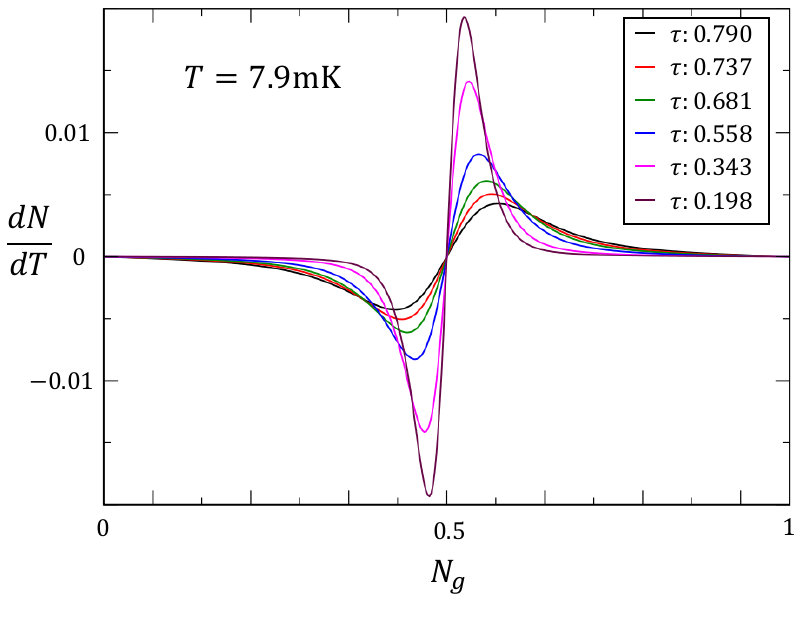}
\caption{$dN/dT$ vs $N_g$ for the 3CK model with different transmissions at $T=7.9$mK. Parameters tuned as in Fig.~2 to match the experimental conductance lineshapes (See Table \ref{tab:param}). Here we measure energy in units of mK.} \label{fig:dndtCB3CK}
\end{figure}

The above model is solved using Wilson's NRG method \cite{wilson1975renormalization,*bulla2008numerical}, which involves the logarithmic discretization of the conduction electron bands,  mapping them to Wilson chains, and then iteratively diagonalizing the discretized model. Retaining only $N_s$ of the lowest energy states at each step amounts to an RG procedure in which the physics on progressively lower energy scales is revealed. 

However, standard NRG cannot be used here due to the complexity of the model at hand (in particular its three conduction channels). Instead, we use the `interleaved NRG' (iNRG) method \cite{mitchell2014generalized,*stadler2016interleaved}, which involves mapping all three channels to a single generalized Wilson chain. This dramatically lowers the computational cost of such calculations, and brings the numerical solution of the 3CK model within reach. We exploit conserved charge in each channel and conserved total spin, use a discretization parameter $\Lambda=3$ and keep $N_s=60000$ states at each step of the calculation.

To compare with experiment, we calculate the dc linear response differential conductance,
\begin{equation}\label{eq:cond}
G_{\alpha\beta}=\frac{dI^{\alpha}}{dV_b^{\beta}}\Bigg|_{V_b^{\beta}\to 0}
\end{equation}
where $I^{\alpha}=-e\langle \dot{N}_{\alpha\uparrow}\rangle$ is the current into lead $\alpha$ and $V_b^{\beta}$ is the voltage bias applied to lead $\beta$. Here $\dot{N}_{\alpha\uparrow}=\tfrac{d}{dt}\hat{N}_{\alpha\uparrow}$ and $\hat{N}_{\alpha\uparrow}=\sum_{k}c_{\alpha k \uparrow}^{\dagger}c_{\alpha k \uparrow}^{\phantom{\dagger}}$. An ac voltage bias on lead $\beta$ can be incorporated by a source term in the Hamiltonian, $H_{\rm bias}^{\beta}=-eV_b^{\beta}\cos(\omega t) \hat{N}_{\beta\uparrow}$, where $\omega$ is the ac driving frequency. The dc limit is obtained as $\omega \to 0$. 

The geometry of the charge Kondo device means that conductance cannot be related simply to a dot spectral function, and we must use the Kubo formula instead,
\begin{equation}\label{eq:kubo}
    G_{\alpha\beta}/G_0 = \lim_{\omega \to 0} \frac{-2\pi  ~{\rm Im}K_{\alpha\beta}(\omega)}{ \omega} \;,
\end{equation}
where $G_0=e^2/h$ as before, and $K_{\alpha\beta}(\omega)=\langle\langle \dot{N}_{\alpha\uparrow} ; \dot{N}_{\beta\uparrow}\rangle\rangle$ is the Fourier transform of the retarded current-current correlator $K_{\alpha\beta}(t)=-i\theta(t)\langle [\dot{N}_{\alpha\uparrow},\dot{N}_{\beta\uparrow}(t) ] \rangle$. Within iNRG, ${\rm Im}K_{\alpha\beta}(\omega)$ may be obtained from its Lehmann representation using the full density matrix technique \cite{weichselbaum2007sum}. In fact, the numerical evaluation is substantially improved by utilizing the identity ${\rm Im}K_{\alpha\beta}(\omega)=-\omega^2 {\rm Im}\langle\langle \hat{N}_{\alpha\uparrow} ; \hat{N}_{\beta\uparrow}\rangle\rangle$ \cite{transport}. 
Due to the channel symmetry considered here, $G_{\alpha\ne\beta}=-2G_{\alpha\alpha}\equiv G$. 

Next, we fine-tuned the model couplings $J$ so as to fit the NRG conductance lineshapes for the Coulomb peak, obtained by varying $N_g$, to the corresponding experimental data at a given transmission $\tau$. Remarkable agreement is obtained over the entire range of $N_g$ for each $\tau$ considered, as shown in Fig.~2(a).
The fitted model $J$ for each experimental $\tau$ is given in Table \ref{tab:param}.

\begin{table*}[t]
\begin{center}
 \begin{tabular}{| c || c | c | c | c | c | c |} 
 \hline
Experimental transmission, $\tau$ & ~0.790~ & ~0.737~ & ~0.681~ & ~0.558~ & ~0.343~ & ~0.198~ \\ 
 \hline
 NRG coupling, $J/D$ & 0.41 & 0.39 & 0.37 & 0.34 & 0.27 & 0.22 \\ 
 \hline
 NRG effective transmission, $\tau_{\rm eff}$~ & 0.83 & 0.79 & 0.75 & 0.68 & 0.51 & 0.37 \\ [1ex] 
 \hline
\end{tabular}
\end{center}
  \caption{Optimized NRG couplings use to fit calculated conductance lineshapes to experimental data in Fig.~2(a).}\label{tab:param}
\end{table*}
  
 The effective QCP transmission $\tau_{\rm eff}$ can be estimated from the low-temperature conductance of a tunnel junction between two non-interacting leads, where the tunneling matrix element is $J$ and all other interactions are ignored. This yields the standard result,
\begin{equation}\label{eq:tau_eff}
    \tau_{\rm eff} = \frac{4(\pi\nu J)^2}{[1+(\pi\nu J)^2]^2}  \;,
\end{equation}
 also given for comparison in Table \ref{tab:param}. Although a good rule of thumb, it is evident that quantitative simulation requires a more sophisticated fitting procedure than simply using $\tau_{\rm eff}$.
 
Importantly, for $\tau=0.79$ and low temperatures $T=7.9$mK, the conductance approaches closely the nontrivial 3CK fixed point value of $G/G_0=2\sin^2(\pi/5)$, as also captured by NRG. We note however that at large transmissions $\tau \sim 0.79$, the Kondo temperature $T_K$ found in the experiment diverges \cite{iftikhar2018tunable}, indicating that a macroscopic number of dot charge states are involved in the screening process. This effect cannot be captured with NRG, which necessitates using a relatively small finite $\bar{N}$ (the maximum $T_K$ in NRG is around $E_C$). This means that the comparison between experiment and theory at high transmission breaks down at elevated temperatures. We therefore confine our discussion here to the experimental base temperature $T=7.9$mK ($\ll E_C$) which still affords a good comparison at large $\tau$ (and noting that such issues do not arise at smaller $\tau$ since then $T_K\ll E_C$ anyway).

Finally, we comment on the static thermodynamic quantity $N=\langle \hat{N}_d\rangle -N_0$, which can be calculated at any temperature $T$ within NRG. For the same set of parameters as used for Fig.~2(a), we obtain $N$ as a function of gate voltage, over the Coulomb oscillation period $0 \le N_g \le 1$ at temperatures $T=7.9$mK and $T'=7.901$mK. Although $\Delta N = N(T)-N(T')$ is very small for this $\Delta T=0.001$mK temperature difference, the NRG calculations are highly accurate and allow for an excellent finite-difference approximation to the derivative $dN/dT\simeq \Delta N/\Delta T$ (the exact derivative can also be obtained by differentiable programming techniques \cite{DP}). The results are presented in Fig.~\ref{fig:dndtCB3CK}.

Note in particular that the full model has exact particle-hole symmetry at all integer and half-integer values of $N_g$. This means that $N$ is a constant and independent of temperature at these points. This is reflected in Fig.~\ref{fig:dndtCB3CK} by the condition $dN/dT=0$ at $N_g=0$ and $N_g=\tfrac{1}{2}$. It is crucial to capture this boundary condition behaviour in the NRG simulations of the model, since $dN/dT$ must be integrated over (half of) the  whole Coulomb oscillation period to correctly obtain the entropy via the Maxwell relation. Hence one must keep multiple charge states in the model ($\bar{N}>1$) to capture the correct gate-periodicity of the Coulomb peaks.

At very low temperatures the Coulomb peaks become very narrow such that the minimum conductance $G_{\rm min}\simeq 0$ in the middle of the valley. Only in this universal regime can the results of the two-charge-state model be used. The experiments are not in this fully  universal regime.

\subsection{Universality of the relation between $dN/dT$ vs $G$ for 3CK}
\begin{figure}[t]
\includegraphics[width=1\columnwidth]{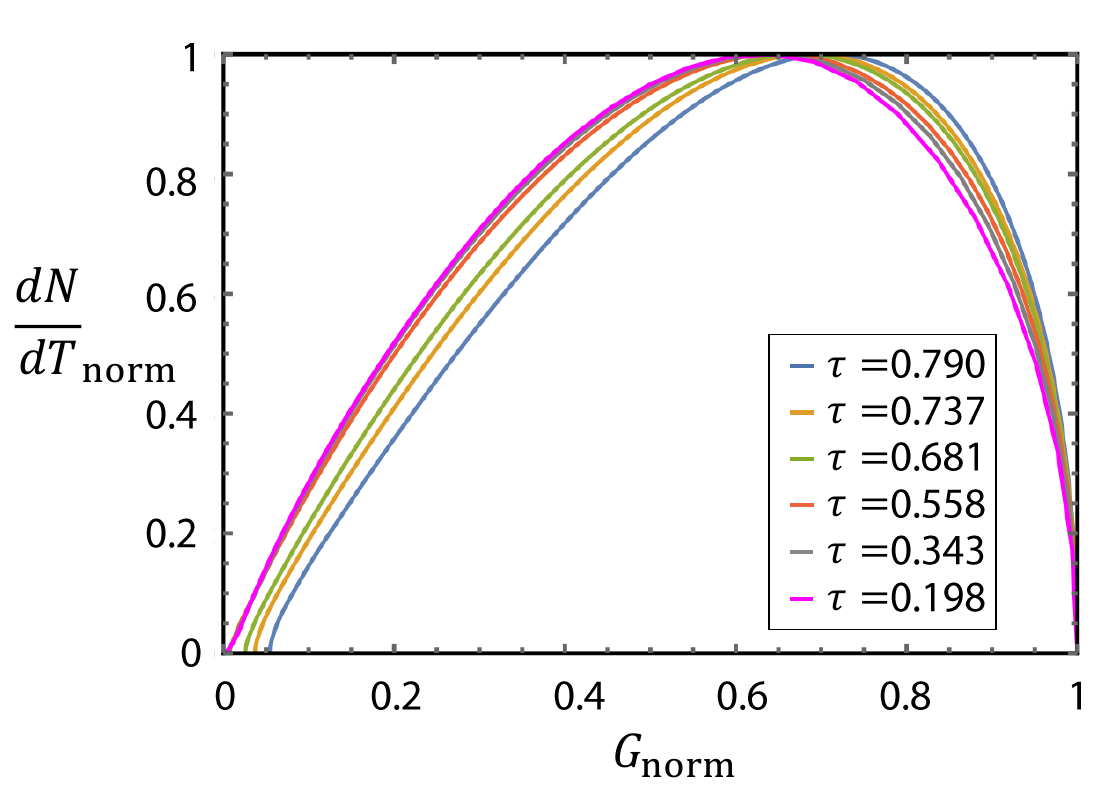}
\caption{Normalized $dN/dT_{\rm norm}$ versus normalized $G_\text{norm}$ is plotted. } \label{fig:3CKuniversal}
\end{figure}
Different from the CB and 2CK cases, there is no simple relation between $dN/dT$ and $G$ for 3CK. 
Figure.~\ref{fig:3CKuniversal} shows normalized $dN/dT$ versus normalized $G$ for various values of $\tau$. The normalized curves in Fig.~\ref{fig:3CKuniversal} do not collapse onto each other. It shows deviations from universality of the relation between $dN/dT$ and $G$, and also indicates that the experimental system is indeed not in the fully universal regime, $T^*\ll T_K$.


\bibliography{QIbasedNonAbelian}